# Chatbot Based Solution for Supporting Software Incident Management Process

Nagib Sabbag Filho*, Rogério Rossi

University of São Paulo, Avenida Prof. Luciano Gualberto, 158, São Paulo, Brazil.

* Corresponding author. Tel.: +55 (11) 99952-6445; email: na.sabbag@gmail.com

**Abstract:** A set of steps for implementing a chatbot, to support decision-making activities in the software incident management process is proposed and discussed in this article. Each step is presented independently of the platform used for the construction of chatbots and are detailed with their respective activities. The proposed steps can be carried out in a continuous and adaptable way, favoring the constant training of a chatbot and allowing the increasingly cohesive interpretation of the intentions of the specialists who work in the Software Incident Management Process. The software incident resolution process accordingly to the ITIL framework, is considered for the experiment. The results of the work present the steps for the chatbot construction, the solution based on DialogFlow platform and some conclusions based on the experiment.

**Key words:** Chatbot platform, decision-making techniques, incident screening, software incident management, software maintenance.

## 1. Introduction

Software Incident Management Process is used to manage the IT incident lifecycle, aimed at IT support professionals and general IT professionals [1]. In order to guarantee the quality of the software in the resolution of the incidents, it is advisable to prepare an Incident Management Plan to support IT organization [2]. The ITIL framework (Information Technology Infrastructure Library), which refers to a framework for efficiency, excellence and consistency in IT service management as proposed by [3] is used to define the main activities related to the Software Incident Resolution Process.

But, how to improve decision making in software incident management process? There are many ways to help decision making, an example to consider is a chatbot, which basically refers to a solution based on NLP (Natural Language Processing) and computing [4]. In general, chatbots can be implemented to contribute to the services of various business areas in organizations of different segments [5]. However, creating a chatbot requires understanding natural language processing as well as mechanisms for knowledge extraction [6].

Is possible to take advantage of many platforms available for chatbots construction, such as Wit.ai, owned by Facebook, Luis, developed by Microsoft, Watson developed by IBM, or Dialogflow provided by Google Inc. [7], [8]. Although platforms have some specific features for chatbot creation, they can all consider an implementation pattern for initial chatbot creation and continuous improvement.

The uniqueness of this research is to use the steps to build a chatbot solution with an existing chatbot platform, allowing to standardize the platform configuration and focus on the conversation flow.

To address the research activities, the results are presented as follows: section two presents the software incidents resolution process; section three introduces some examples of chatbot platforms; section four discusses the steps for implementing a solution based on chatbot for software incident management; and, finally, section five presents conclusive results of the research.

## 2. Software Incident Resolution Process

Software incident occur when an error manifests itself as a malfunction in the software-based system [9]. These errors can be adapted according to the activities defined by the ITIL framework, for example. In this article, the Software Incident Resolution Process to assist support staff in incident correction actions is presented as part of the experiment. In general, the incident resolution activity can be defined as a validation step of the Software Incident Resolution Process, to assess whether the correction has been performed [10]. Validation activities can be performed by both: a support analyst, and the end user; depending on what is defined in the Software Incident Management Process.

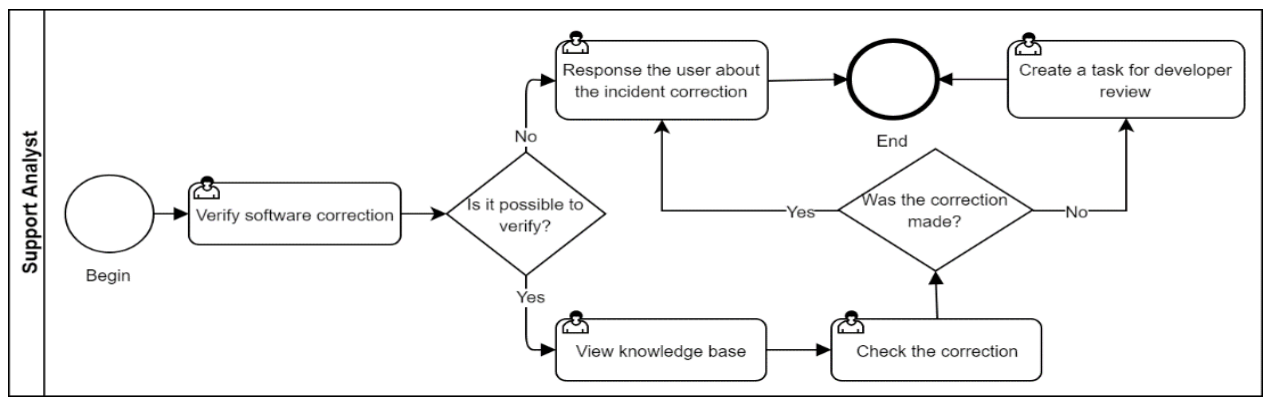

Fig. 1. Software incident resolution process.

Fig. 1 shows the activities of the Software Incident Resolution Process and presents some options for the support analyst to evaluate the corrections of the software incidents. These instructions could be performed through a chatbot, since it can map user actions interacting with a knowledge base to obtain the most appropriate responses [11].

## 3. Examples of Chatbot Platform

Decision-making activities can be executed for a variety of business problems, including for Software Incident Management Process. This type of solutions can be presented through a chatbot based on user experience. Such solutions, henceforth called chatbot (chatting agent that interacts with users using natural language processing [12]) can support the activities of a business area and its decision-making actions. The development of a chatbot solution is based on the programming of intentions where they create several questions and a set of answers for each question [13]. The general characteristics of three platforms for the implementation of chatbot solutions are: 1) Watson from IBM; 2) DialogFlow by Google Inc.; and 3) Luis from Microsoft.

Watson offers the ability to process natural language through a series of linguistic analysis (for example morpholexical and syntactic analysis) and its generation of hypotheses to identify a response in its knowledge base makes this platform very popular for the creation of chatbots [14].

DialogFlow platform also used for chatbots development, modeling chatbot activities using concepts such as intentions, entities and contexts [15]. The platform also allows training actions for the chatbot to learn with intentions not yet understood.

Luis platform also uses the classification of intentions and entities, as well as the creation of dialogues for each intention. The Luis platform presents an obligation in the definition of values for the entities before the recognition of them, since it can have a behavior of extracting entities dynamically [16].

All three platforms are relevant for creating chatbots. In this article, DialogFlow is considered for the implementation of a chatbot solution to aid in the decisions of the software support team regarding the software incident resolutions.

## 4. Implementation of a Chatbot Based Solution Using DialogFlow Platform

The steps for implementing solutions based on chatbots platforms are described below in order of execution and can be adapted to favor other scenarios or platforms for chatbots. The main objective of these steps is to facilitate the implementation of this type of solution and to favor the continuous evolution of chatbot learning. The steps can be performed continuously, favoring the iterative and incremental development of the solution.

### 4.1. Collection of Software Incidents Samples

The collection of types of incidents for software products is carried out considering as much information as possible to supervise chatbot learning, using supervised learning methods [17]. Table 1 presents some types of software incidents, according to their functionalities.

Table 1. Samples of Types of Software Incidents

| Sample | Description |
| --- | --- |
| Software Unavailable | Unavailability or slowness when using a software feature |
| Software access failure | Inability to access the system |
| Incomplete report | Report with no data |
| Data import failed | Data import agents do not run at the specified time |
| Incorrect software calculations | Equations that compromise the outcome of reports |

Still in this step, the samples should be transformed into actions that are performed by the support professional, when acting in a software incident, to validate if the correction was performed. Fig. 2 presents these actions drawn in a decision tree [17].

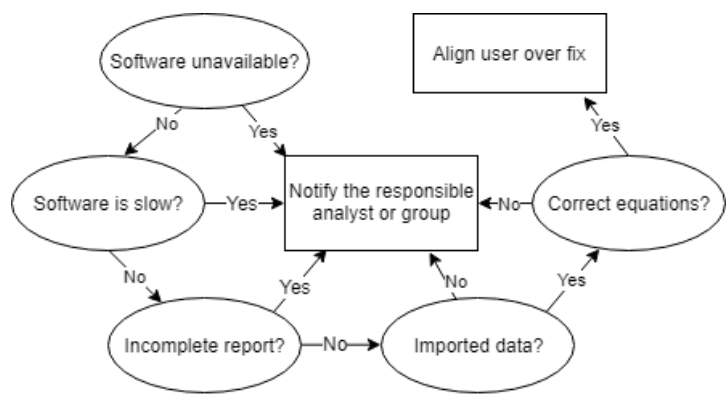

Fig. 2. Actions to validate software adjustments.

It is possible that these actions differ among the various types of software product. This experiment consists in performing the actions of Fig. 2 to verify if the correction was performed by the responsible analyst or group, but the actions are adaptable according to the need of each software product.

### 4.2. Importing Intentions into Chatbot

In this step, intentions are presented to identify the actions that must be followed according to the understanding of the natural language that was obtained in the user's message [18]. Accordingly, to the decision tree presented in Fig. 2, intentions can be created as shown in Table 2.

Table 2. Intentions

| Intention | Response | Condition | Action |
|---|---|---|---|
| Intention 01 | Software Incident? | Yes | Proceed for intention 02 |
| Intention 02 | Is the Software unavailable? | Yes | Proceed for intention 08 |
| Intention 02 | Is the Software unavailable? | No | Proceed for intention 03 |
| Intention 03 | Is the software slow? | Yes | Proceed for intention 08 |
| Intention 03 | Is the software slow? | No | Proceed for intention 04 |
| Intention 04 | Is the report incomplete? | Yes | Proceed for intention 08 |
| Intention 04 | Is the report incomplete? | No | Proceed for intention 05 |
| Intention 05 | Is the data imported? | Yes | Proceed for intention 06 |
| Intention 05 | Is the data imported? | No | Proceed for intention 08 |
| Intention 06 | The calculus is correct? | Yes | Proceed for intention 07 |
| Intention 06 | The calculus is correct? | No | Proceed for intention 08 |
| Intention 07 | Align user over fix | | |
| Intention 08 | Notify the responsible analyst or group | | |

In Table 2, the 'intention' column serves as an identifier. The 'response' column represents the chatbot response for a given user message [19]. The 'condition' column serves as a guideline to proceed as a given action, this same column can represent several training phrases that are a set of words that users can inform chatbot [19]. Finally, the 'action' column represents the continuation of the intention according to the condition that was understood.

### 4.3. Monitoring the Processing of Intentions

After setting up the intentions, the monitoring process of the chatbot is performed to verify if it is triggering the correct intentions according to the requests that are received. The Dialogflow platform considers the fallback action every time the user message does not trigger any intention [20]. This is useful for chatbot learning with calls that were not understood and improving intentional training phrases.

By monitoring and refining intentions, the cycle continues to return to the first step to identify new examples of actions being taken by support analysts. In this way the stages characterize a cycle of continuous improvement of the chatbot, being carried out in a constant and uninterrupted way.

### 5. Conclusion

Considering the steps to implement solutions based on chatbots platforms, it is possible to delegate the responsibility of collecting samples of intentions to any professional of support team, not having to be an expert in chatbot. The formalization of the examples also allows the change of chatbot platform without losing the intentions actions, contributing with the selection of the most relevant platform for each scenario. The creation of the table of intentions allows the professional to focus on the conditions and not the training phrases (leaving this responsibility to the platform expert to create, according to the understanding of the natural language of each platform). The collection of samples of software incidents enables those interested to understand the origin of the actions that were explicit in chatbot. Finally, monitoring allows continuous engagement with solution improvement and supports the team with chatbot learning, through the rescue of intentions that were not understood.

One of the limitations that can be observed based on this work is its use in areas where the process is not well defined. The lack of a mapped and defined process compromises the creation of the dialogue flow, especially in collecting examples from the first step. For proper implementations of chatbots, is necessary to verify that the area makes use of best practices of service management process. One restriction that may be considered is the limitation of chatbot platforms, which may cause a reevaluation of the intent import step.

Given these results, it can be concluded that these steps contribute significantly to the implementation of a chatbot based solution. One possible future research regarding this theme would be to follow the steps using another platform or processes related to the Software Incident Management Process.

## Conflict of Interest

The authors declare no conflict of interest.

## Author Contributions

Nagib Sabbag Filho conducted the primary research and drafted the article. Rogério Rossi provided active feedback on the drafts for revision. All authors had approved the final version.

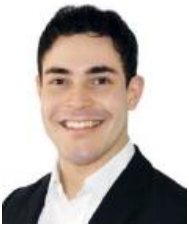
**Nagib Sabbag Filho** was born in São Paulo city, São Paulo, Brazil, in 1989. Nagib received his graduate studies in IT management and governance from Senac University Center, São Paulo, Brazil in 2016. He received his master of business administration (MBA) degree in software technology from University of São Paulo, São Paulo, Brazil in 2019. He completed an MIT Sloan Executive Education Program about Internet of Things. His research interests include software engineering and artificial intelligence.

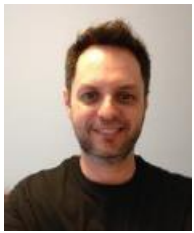
**Rogério Rossi** has a PhD and MSc in computer engineering, both by Mackenzie Presbyterian University as he also has a bachelor's degree in mathematics by the University Center Foundation Santo André. He completed his postdoctoral at the University of São Paulo. He is a professor of software technology and internet of things & data analytics. His currently field of study is related to data analytics, data science and big data also considering academic and learning analytics.